\documentclass{article}
\usepackage{graphicx,subfigure}
\usepackage[square,comma,sort&compress]{natbib}
\begin{document}

\title{A Poisson model \\ for earthquake frequency uncertainties \\ in seismic hazard analysis}

\author{J. Greenhough \& I. G. Main \\ School of Geosciences, University of Edinburgh, UK \\ john.greenhough@ed.ac.uk, ian.main@ed.ac.uk}
\maketitle
\begin{abstract}
Frequency-magnitude distributions, and their associated uncertainties, are of key importance in statistical seismology. When fitting these distributions, the assumption of Gaussian residuals is invalid since event numbers are both discrete and of unequal variance. In general, the observed number in any given magnitude range is described by a binomial distribution which, given a large total number of events of all magnitudes, approximates to a Poisson distribution for a sufficiently small probability associated with that range. In this paper, we examine four earthquake catalogues: New Zealand (Institute of Geological and Nuclear Sciences), Southern California (Southern California Earthquake Center), the Preliminary Determination of Epicentres and the Harvard Centroid Moment Tensor (both held by the United States Geological Survey). Using independent Poisson distributions to model the observations, we demonstrate a simple way of estimating the uncertainty on the total number of events occurring in a fixed time period.     
\end{abstract}

\section{Introduction}\label{intro}

It is well documented that typical catalogues containing large numbers of earthquake magnitudes are closely approximated by power-law or gamma frequency distributions \citep{richter,turcotte,main1,main2}. This paper addresses the characterisation of counting errors (that is, the uncertainties in histogram frequencies) required when fitting such a distribution via the maximum likelihood method, rather than the choice of model itself (for which see \citep{leonard}). We follow this with an empirical demonstration of the Poisson approximation for total event-rate uncertainty \citep[used in][]{leonard}. Our analysis provides evidence to support the assumption in seismic hazard assessment that earthquakes are Poisson processes \citep{reiter,boz,lombardi,koss}, which is routinely stated yet seldom tested or used as a constraint when fitting frequency-magnitude distributions. Use is made of the Statistical Seismology Library \citep{sslib}, specifically the data downloaded from the New Zealand Institute of Geological and Nuclear Sciences (GNS, http://www.gns.cri.nz), the Southern California Earthquake Center (SCEC, http://www.scec.org) and the United States Geological Survey (USGS, http://www.usgs.gov), along with associated R functions for extracting the data. 

Consider a large sample of $N$ earthquakes. In order to estimate the underlying proportions of different magnitudes, which reflect physical properties of the system, the data are binned into $m$ magnitude ranges containing {\bf n} events such that $\sum_{i=1}^{m}n_{i}=N$. Since {\bf n} are discrete, a Gaussian model for each $n_{i}$ is inappropriate and may introduce significant biases in parameter estimations \citep{aki,keilis,sandri}. Hence when fitting some relationship with magnitudes {\bf M}, {\bf n}$_{fit}=f(${\bf M}$)$, linear regression must take the generalised, rather than least-squares, form \citep{mccullagh}. Weighted least squares is an alternative approach which we do not consider here. The set {\bf n} is described as a multinomial distribution; should we wish to test whether two different samples {\bf n} and {\bf n}$'$ are significantly different given a fixed $N$ ``trials'', confidence intervals that reflect the simultaneous occurrence of all {\bf n} must be constructed using a Bayesian approach \citep{vermeesch}. However, in the case of earthquake catalogues, it is the temporal duration rather than the number of events that is fixed. Observational variability is not, therefore, constrained to balance a higher $n_{i}$ at some magnitude with a lower $n_{j}$ elsewhere, and {\bf n} are well approximated by independent binomial distributions \citep{johnson}.

Each incremental magnitude range $(M_{i}-\delta M/2$,$M_{i}+\delta M/2)$ contains a proportion of the total number of events and hence a probability $p_{i}$ with which any event will fall in that range. Providing the overall duration of the catalogue is greater than that of any significant correlations between either magnitudes or inter-event times, $n_{i}$ can be modelled as a binomial experiment with $N$ independent trials each having a probability of ``success'' $p_{i}$ \citep{johnson}. The binomial distribution converges towards the Poisson distribution as $N\rightarrow\infty$ while $Np_{i}$ remains fixed. Various rules of thumb are quoted to suggest values of $N$ and $p_{i}$ for which a Poisson approximation may be valid; see for example \citep{greenturn,borradaile}. Here, we show empirically in Sect.~\ref{fmd} that the frequencies in four natural earthquake catalogues are consistent with a Poisson hypothesis, while in Sect.~\ref{eru} we derive the resulting Poisson distributions of the total numbers of events, which provide simple measures of uncertainty in event rates.  

\section{Frequency-magnitude Distributions}\label{fmd}

Four earthquake catalogues are analysed: New Zealand (1460 -- Mar 2007), Southern California (Jan 1932 -- May 2007), the Preliminary Determination of Epicentres (PDE, Jan 1964 -- Sep 2006) and the Harvard Centroid Moment Tensor (CMT, Jan 1977 -- June 1999, $<$100 km focal depth). While we impose no additional temporal or spatial filters on the raw data, magnitude limits are chosen to minimise the effects of incompleteness at lower magnitudes and undersampling of higher magnitudes. Following \citep{leonard}, who demonstrate the use of an objective Bayesian information criterion for choosing between functions, we seek to fit each catalogue with either a single power-law distribution 
\begin{equation}
\log_{10}{\bf n}=a-b{\bf M} \label{eqplaw},
\end{equation} 
{\bf M} being already on a log scale, or a gamma distribution
\begin{equation}
\log_{10}{\bf n}=a-b{\bf M}-c\exp(k{\bf M}) \label{eqgam},
\end{equation}
where a, b, c and k are constants. The gamma distribution consists of a power law of seismic moment or energy at the lower magnitudes followed by an exponential roll-off. Unlike pure power laws, its integration is finite and so it represents a physical generalisation of the Gutenberg-Richter law; for examples see \citep{koravos} and references therein. For internal consistency, the Poisson assumption in \citep{leonard} is indeed valid as we now demonstrate.

\begin{figure}
\centering
\subfigure[]{\includegraphics[width=0.45\textwidth]{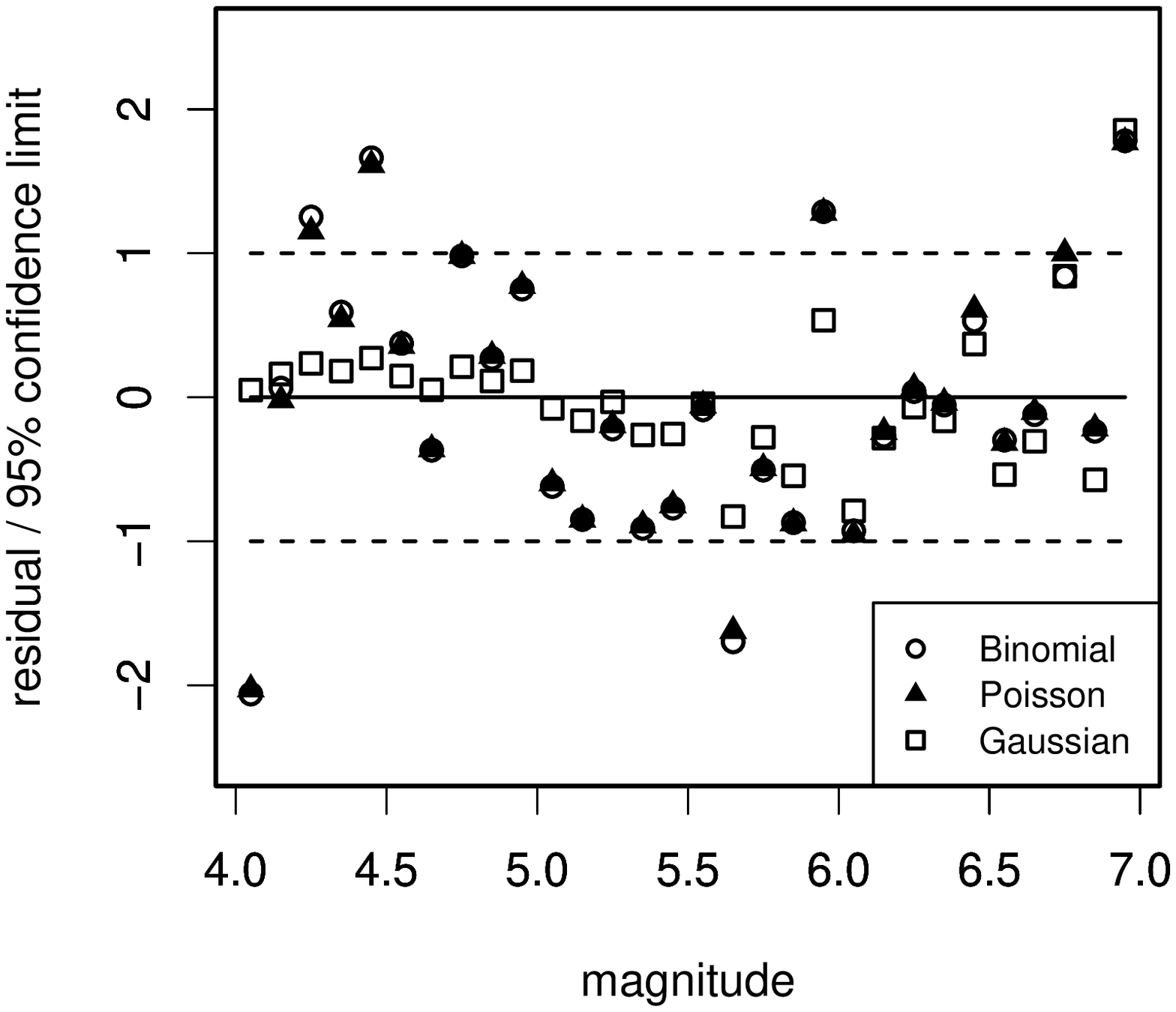}}
\subfigure[]{\includegraphics[width=0.45\textwidth]{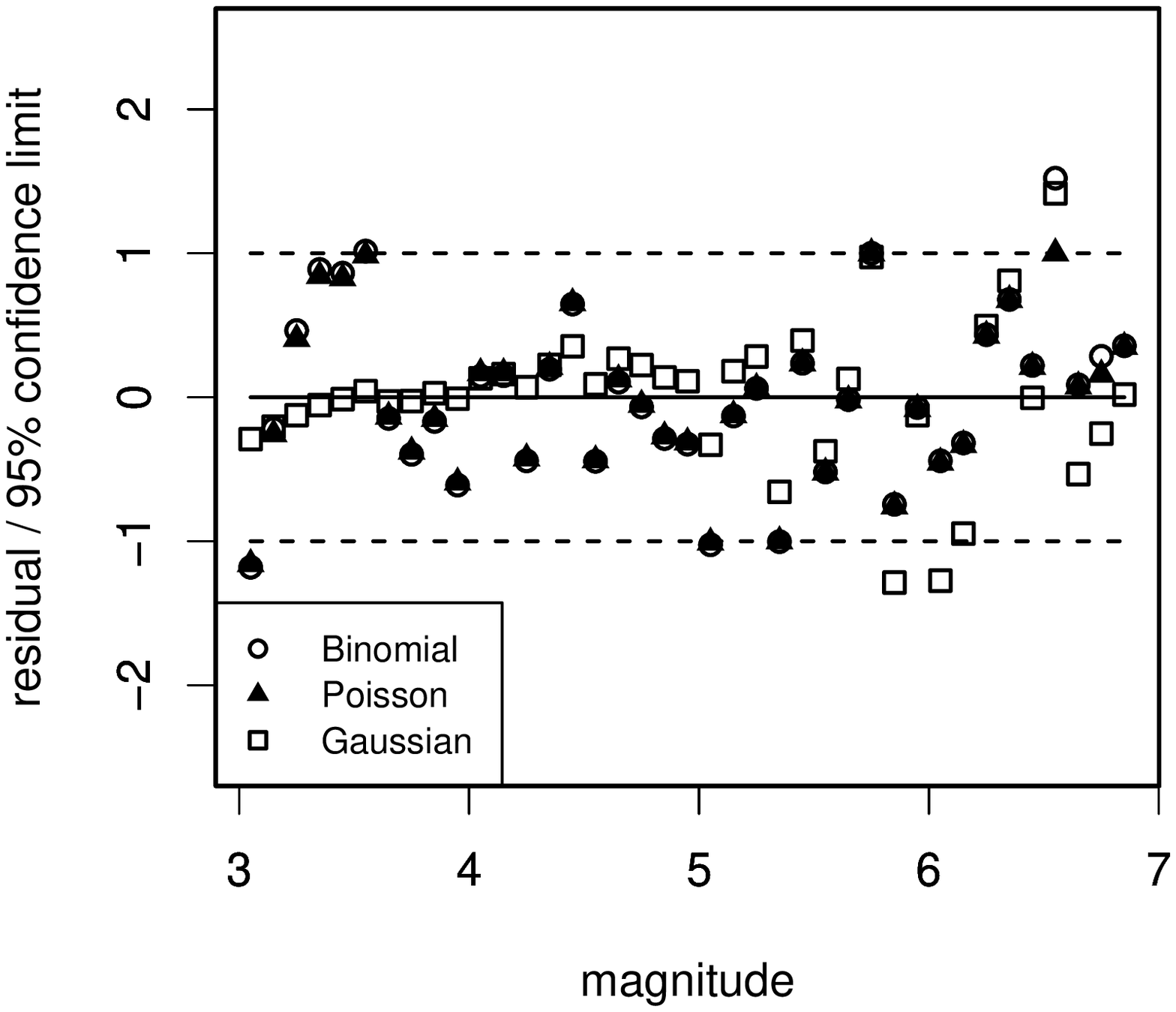}}
\subfigure[]{\includegraphics[width=0.45\textwidth]{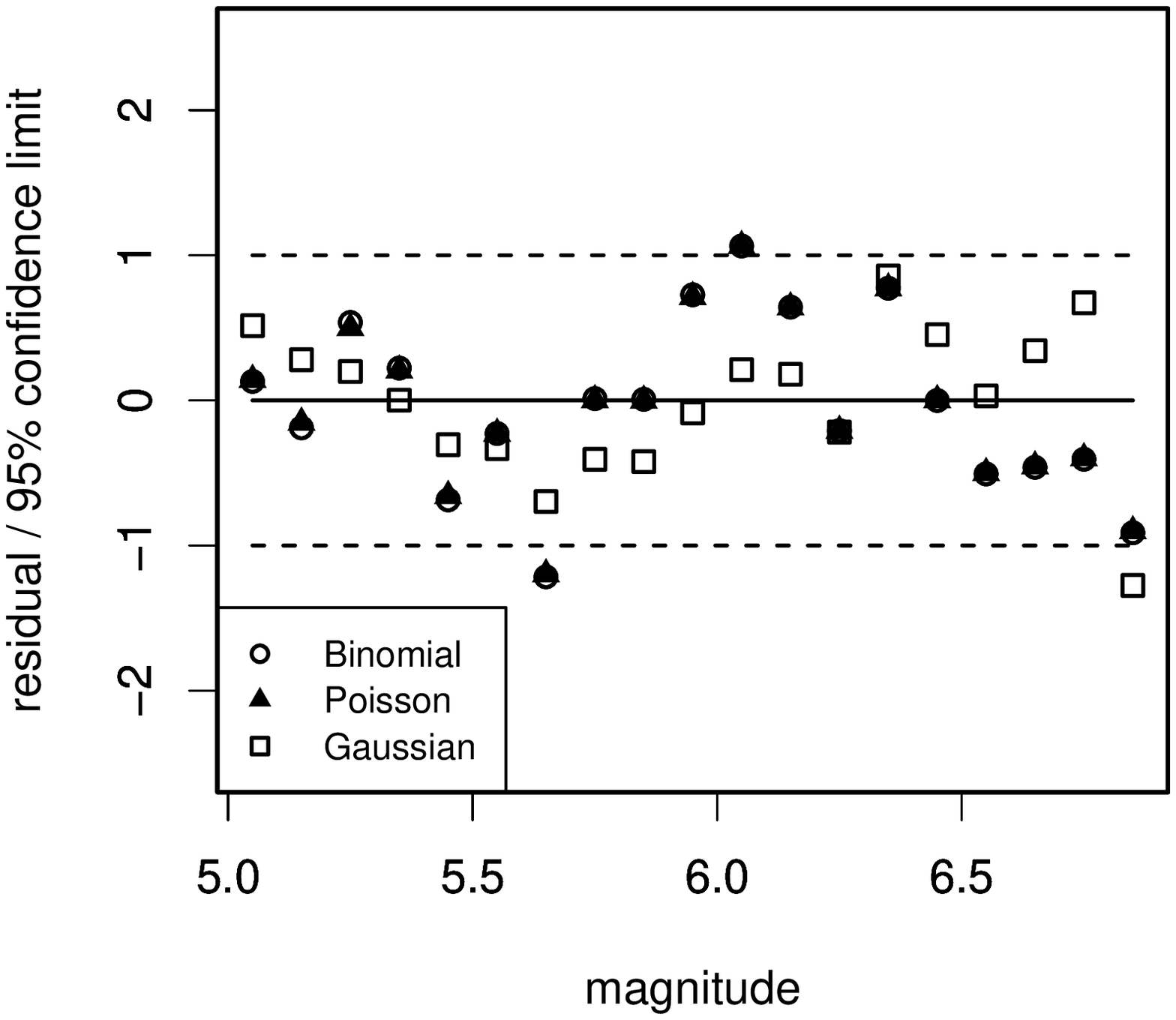}}
\subfigure[]{\includegraphics[width=0.45\textwidth]{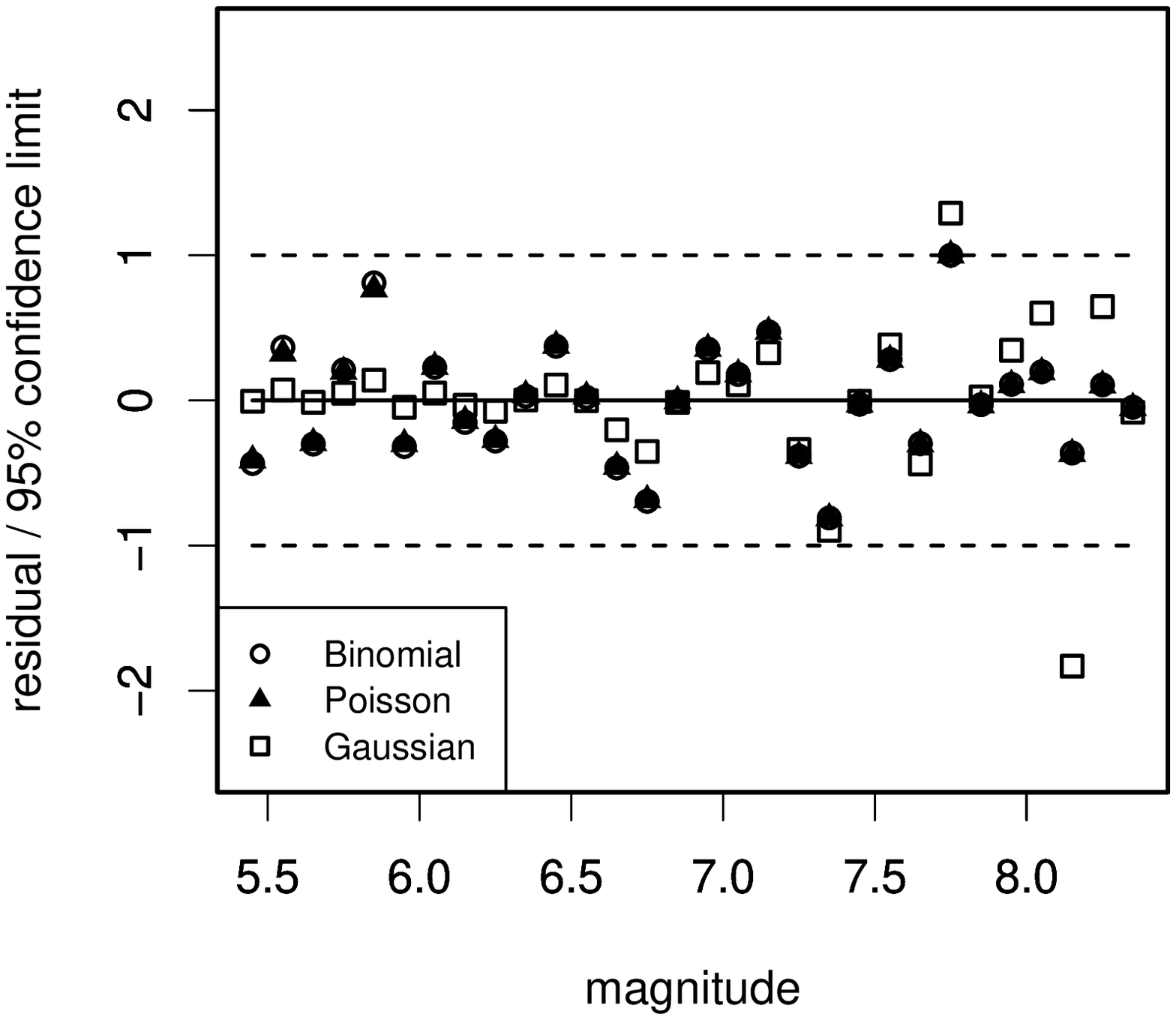}}
\caption{Residuals of fitted frequency-magnitude distributions from GNS/SCEC/USGS catalogues: (a) New Zealand, (b) Southern California, (c) PDE, (d) CMT. (solid line) Best fit to Eq.~\ref{eqplaw} or \ref{eqgam}; (dashed lines) 95\% confidence limits of respective distribution.}
\label{resid}
\end{figure}

\begin{figure}
\centering
\subfigure[]{\includegraphics[width=0.45\textwidth]{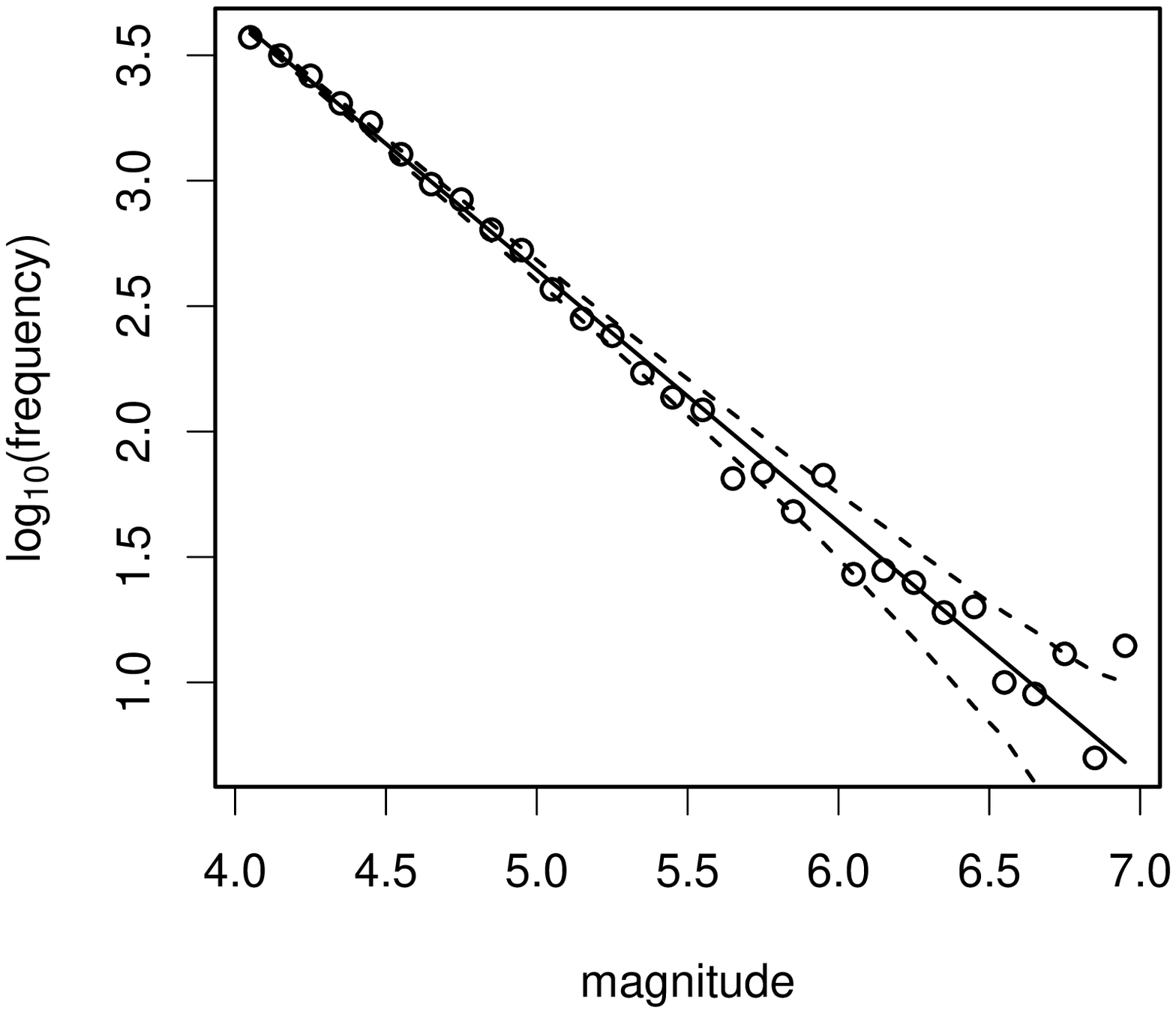}}
\subfigure[]{\includegraphics[width=0.45\textwidth]{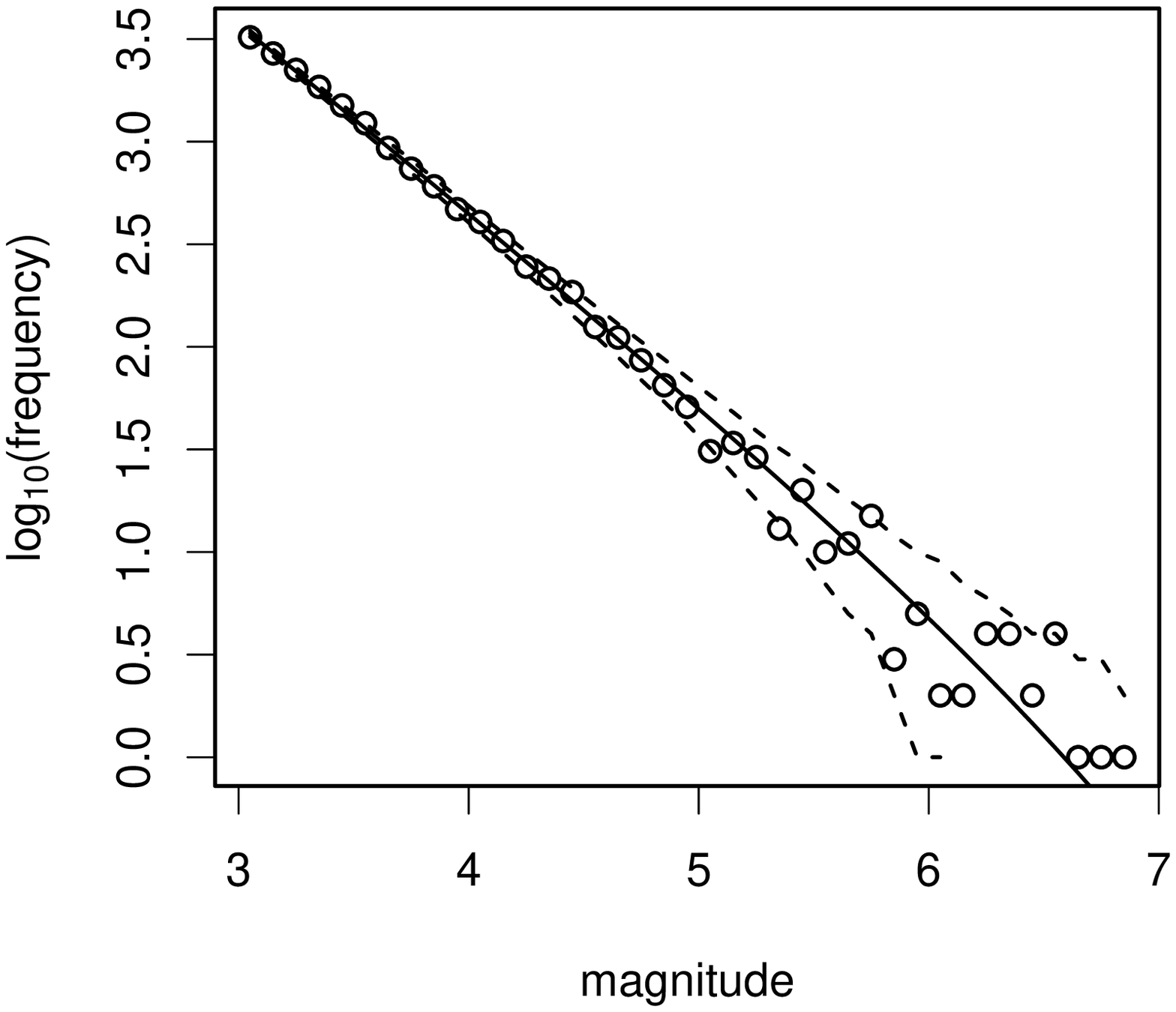}}
\subfigure[]{\includegraphics[width=0.45\textwidth]{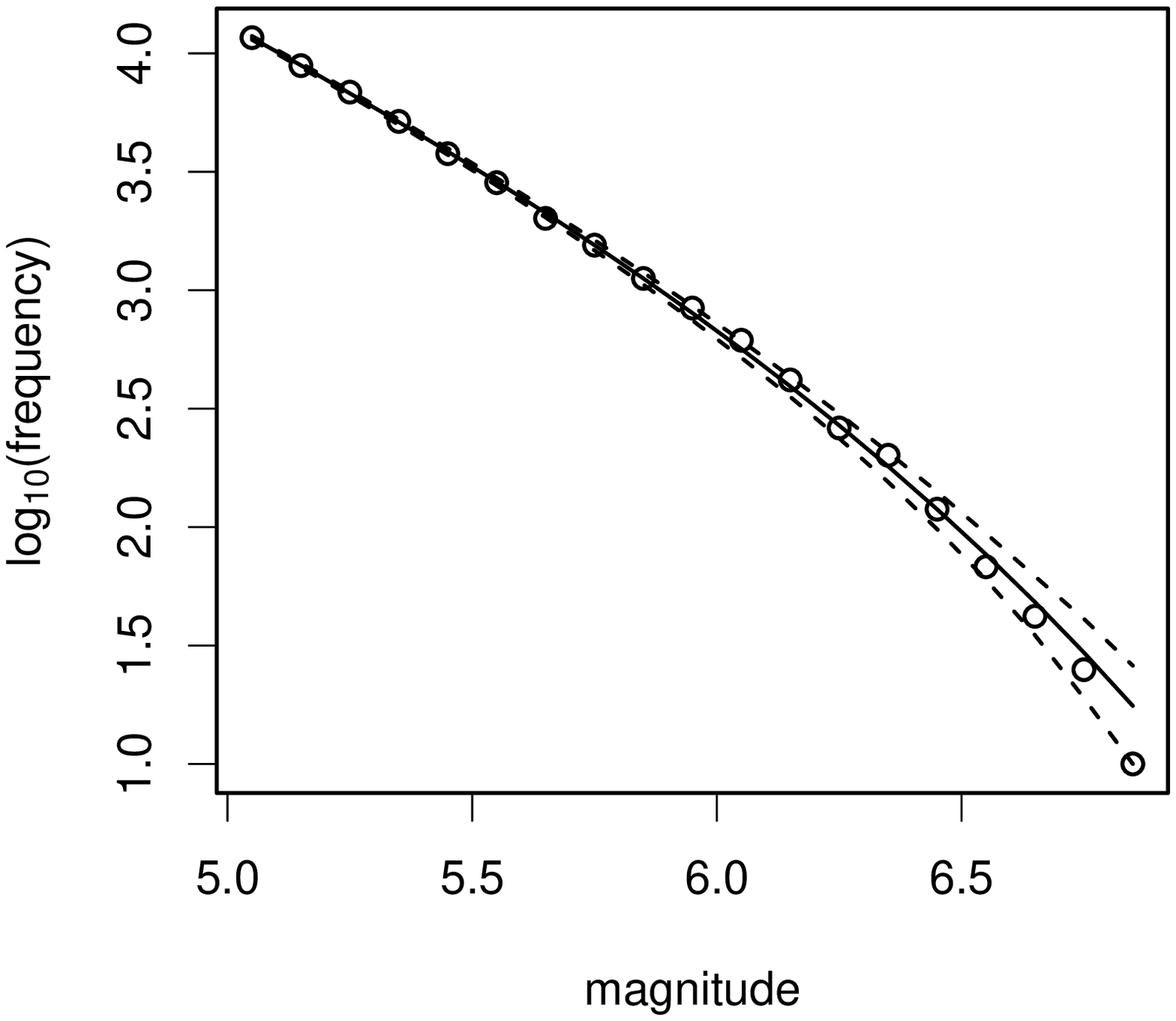}}
\subfigure[]{\includegraphics[width=0.45\textwidth]{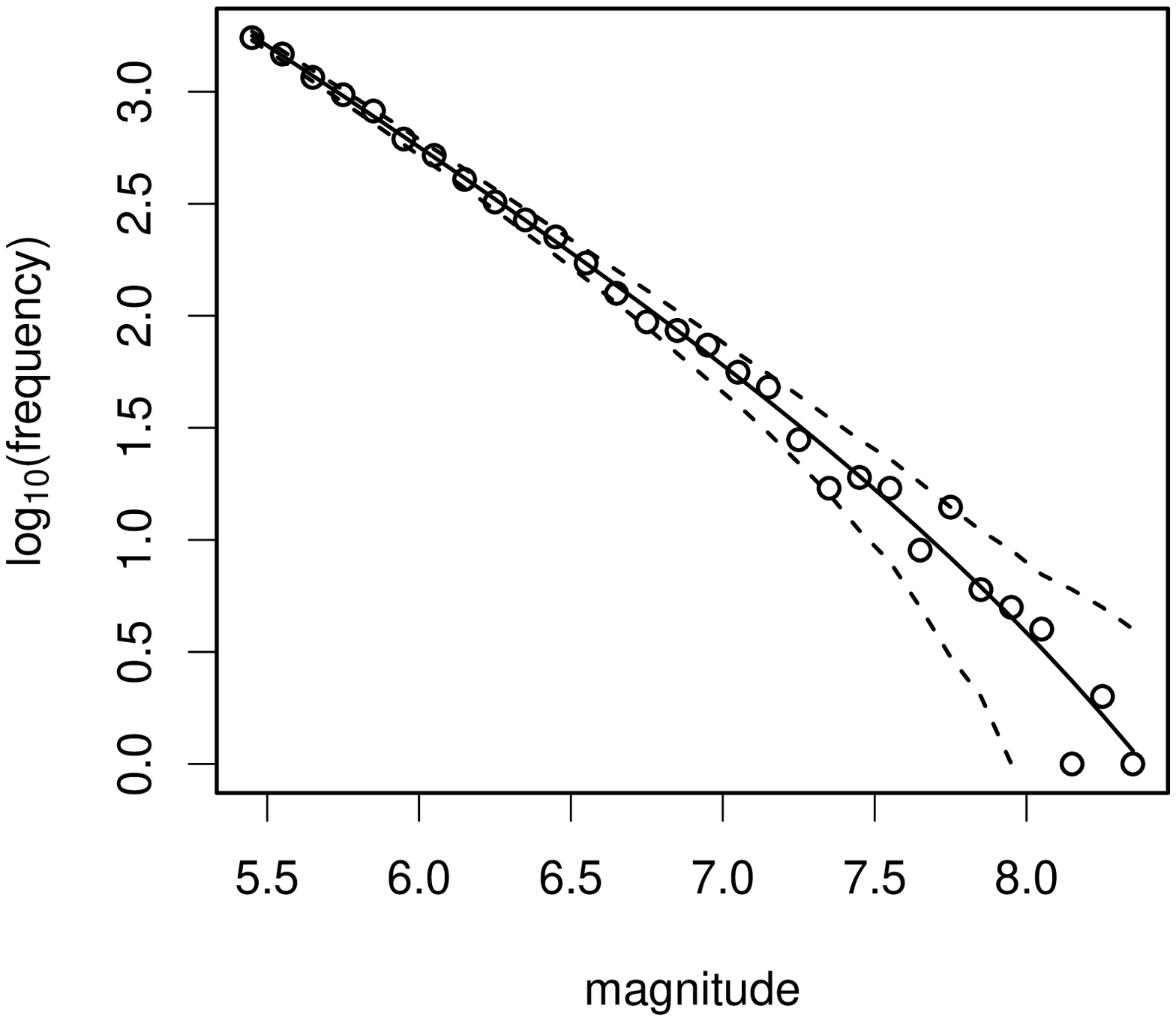}}
\caption{Frequency-magnitude distributions from GNS/SCEC/USGS catalogues. (solid line) Best fit to Eq.~\ref{eqplaw} or \ref{eqgam}: (a) New Zealand, power law $b=1.0$; (b) Southern California, gamma $b=0.91$; (c) PDE, gamma $b=0.91$; (d) CMT, gamma $b=0.85$. (dashed lines) 95\% Poisson confidence limits. Unweighted Gaussian regression leads to $b$-value estimates of (a) 0.98, (b) 1.03, (c) 0.66, (d) 0.83.}
\label{fits}
\end{figure}

As explained in Sect.~\ref{intro}, generalised linear regression is required since we have non-Gaussian counting errors on each bin. To test the consistency of these counting errors with the Gaussian, binomial and Poisson distributions, the residuals (observations minus chosen fit) are normalised to their 95\% confidence intervals and plotted in Fig.~\ref{resid}. 
In all four catalogues, the binomial and Poisson residuals are almost indistinguishable, and show no significant deviation from the expected 1 in 20 exceedance rate when counting those points that lie outside the 95\% confidence limits. Equal bin widths $\Delta M=0.1$ are used as is common practice in earthquake hazard analysis; while this underestimates the intrinsic physical uncertainty of earthquake magnitude determination, for the present purposes the Poisson model appears to be a good proxy. At least, for the catalogues considered here and with $\Delta M=0.1$, the Poisson model is valid. By way of a further check, the value $b$ of the fitted power-law slope (Equs.~\ref{eqplaw}, \ref{eqgam}) given binomial errors is, to two significant figures, equal to that given Poisson errors, for all four catalogues. Constant Gaussian errors systematically overestimate frequency uncertainties on the smaller magnitudes, leading to differences in $b$ of $+10\%$ and $-30\%$ respectively for the Southern California and PDE data (see caption of Fig.~\ref{fits}). These are caused by over-weighting the exponential components of the gamma distributions and exemplify worst-case results of incorrect error structures. In Fig.~\ref{fits}, then, we need only plot the fits and uncertainties using the Poisson model.
Let us now describe, in Sect.~\ref{eru}, the usefulness of this result for estimating event-rate uncertainties. 

\section{Event-rate Uncertainties}\label{eru}

\begin{figure}
\centering
\subfigure[]{\includegraphics[width=0.45\textwidth]{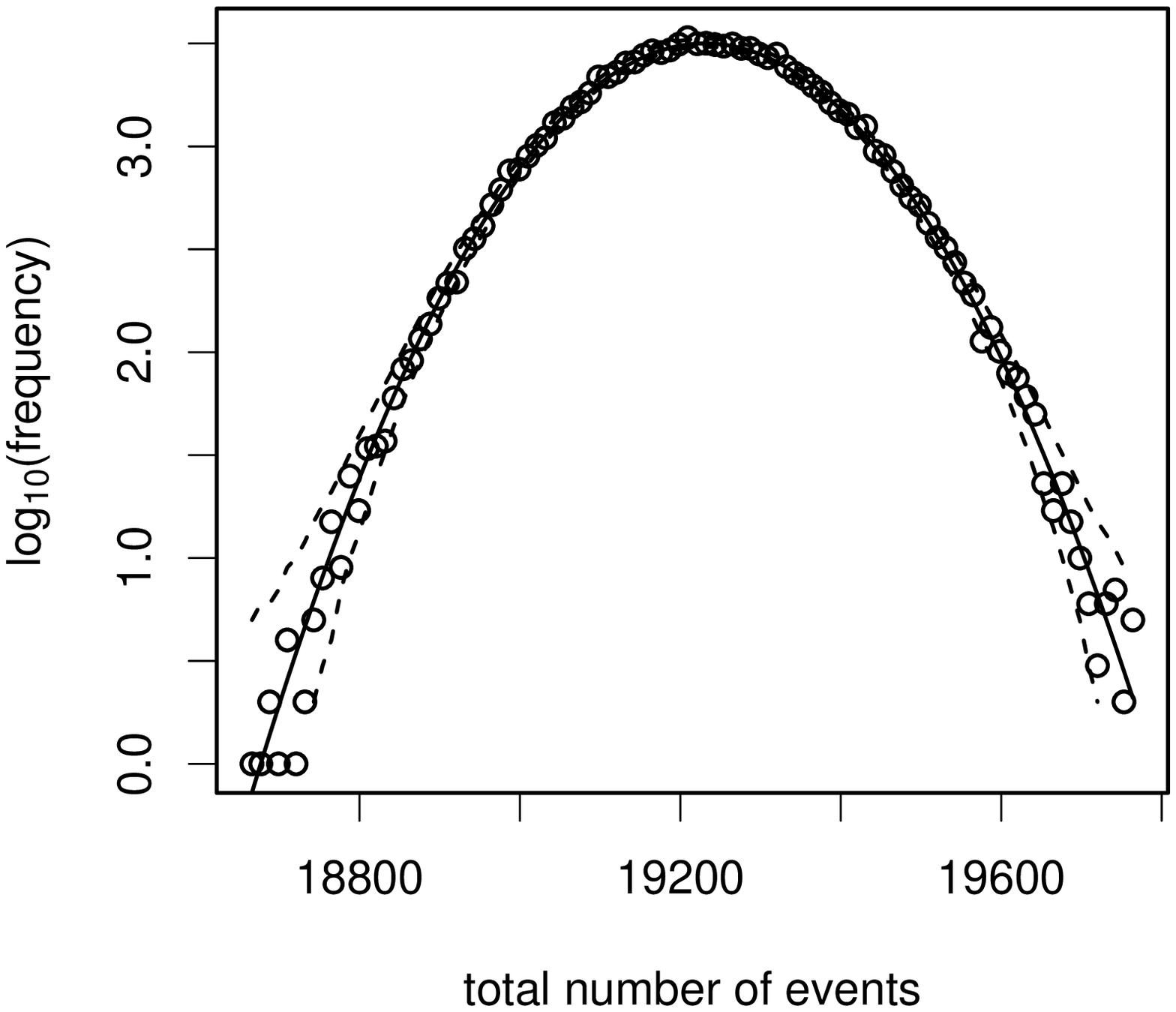}}
\subfigure[]{\includegraphics[width=0.45\textwidth]{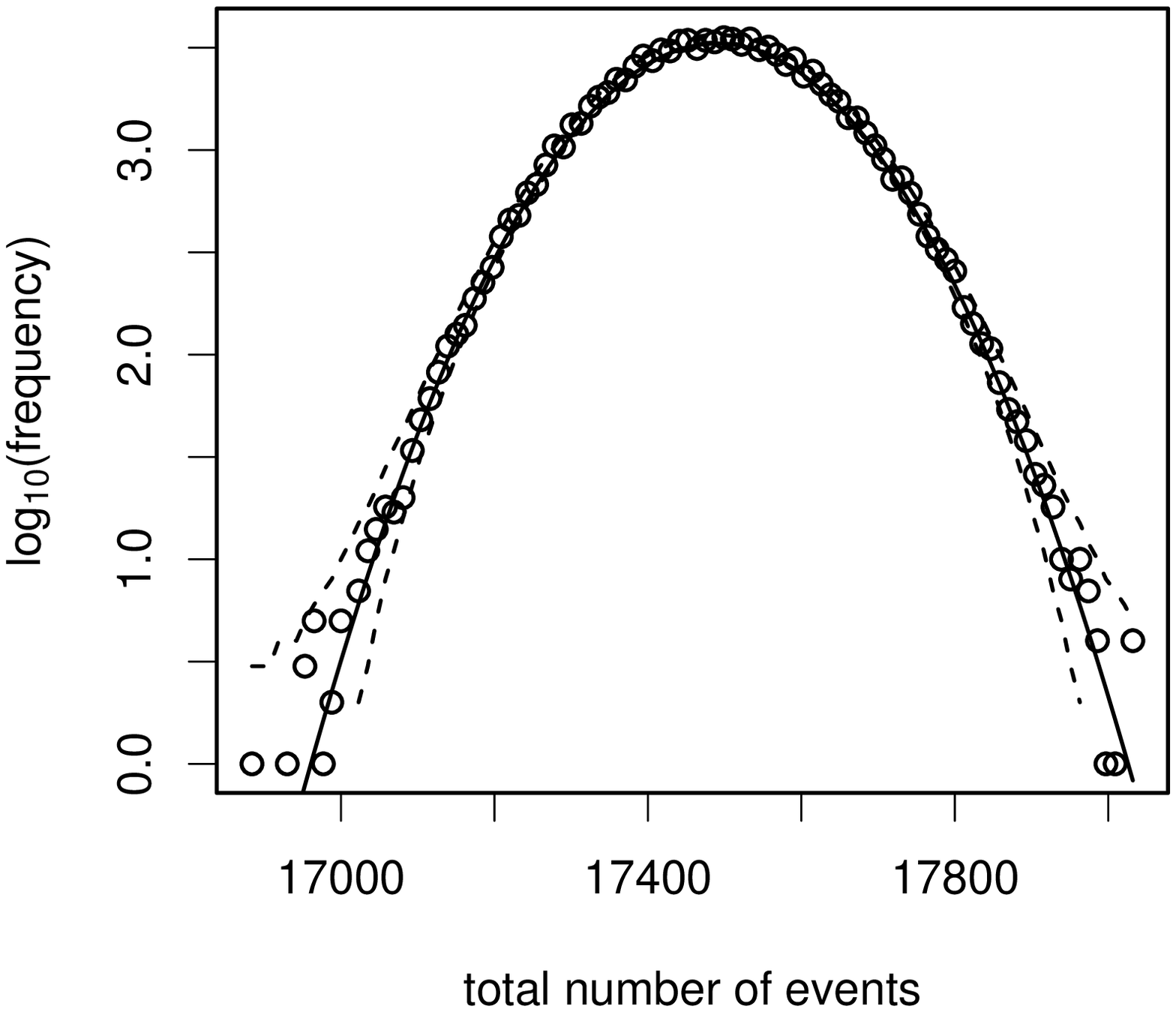}}
\subfigure[]{\includegraphics[width=0.45\textwidth]{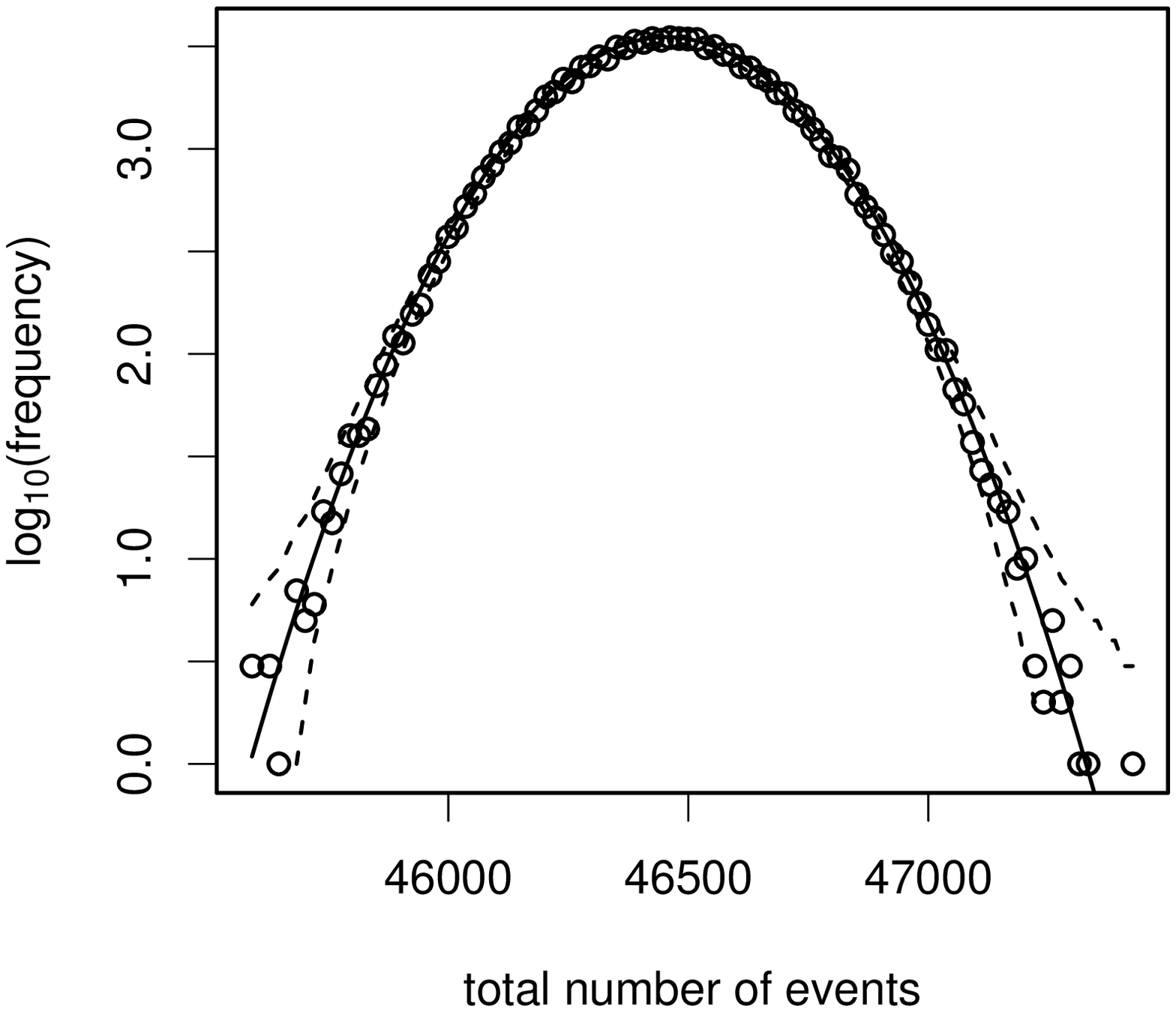}}
\subfigure[]{\includegraphics[width=0.45\textwidth]{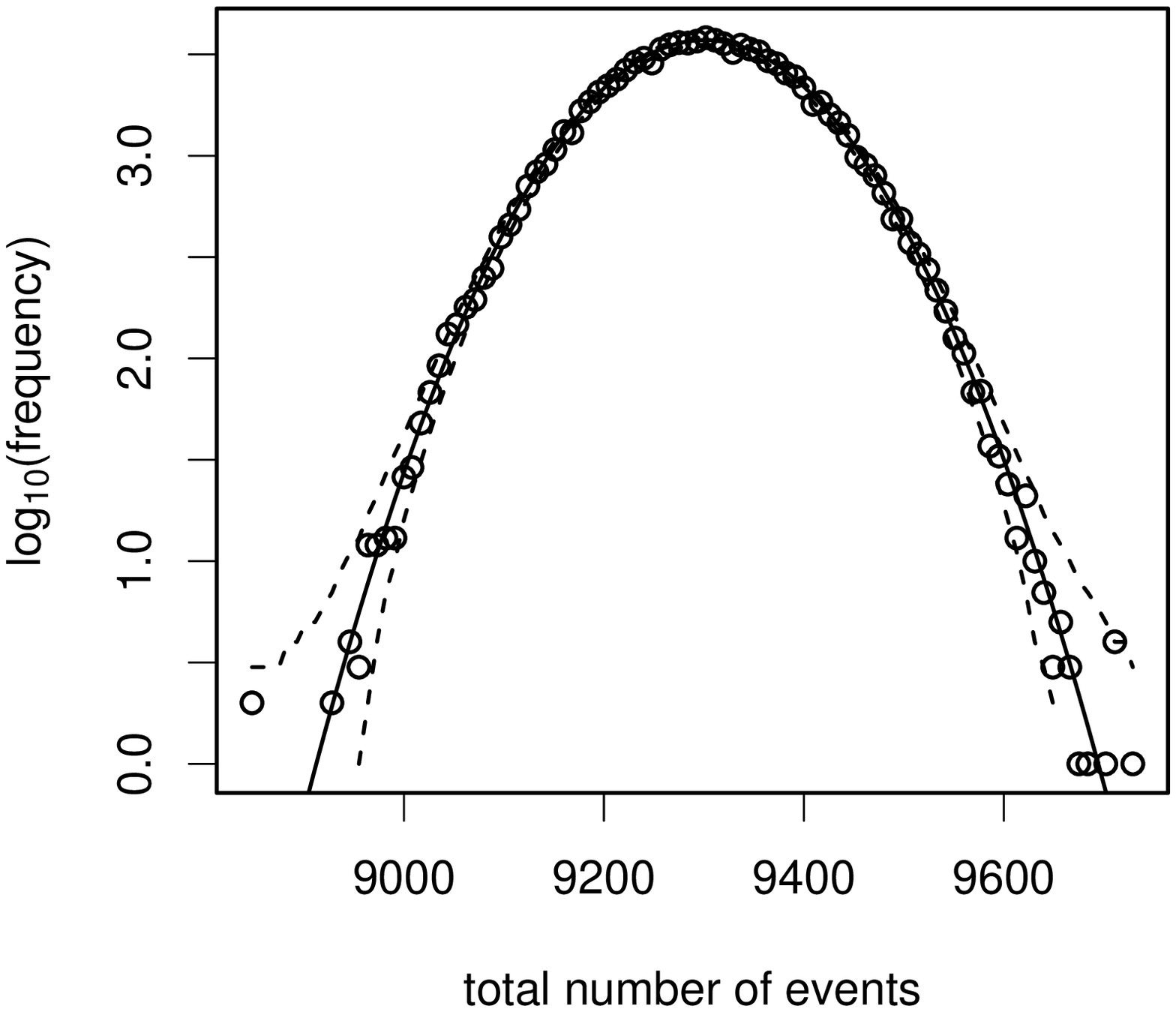}}
\caption{Event-rate distributions from $10^{5}$ simulated realisations of GNS/SCEC/USGS catalogues. Each total event-rate is the sum of a random sample of frequencies, one per bin, given Poisson uncertainties shown in Fig.~\ref{fits}. (a) New Zealand, (b) Southern California, (c) PDE, (d) CMT. (solid line) Best fit Poisson distribution; (dashed lines) 99\% binomial confidence limits.}
\label{rates}
\end{figure}

Having established that independent Poisson distributions characterise the magnitude frequencies in these four catalogues (importantly, these data span sufficiently large times and distances as to minimise dependencies due to clustering), we now ask how this impacts on uncertainties in total numbers of events. While we cannot create equivalent catalogues by re-sampling the same regions under the same physical conditions, we can simulate $S=10^{5}$ samples from each magnitude range by keeping the fitted mean $\lambda_{i}$ constant (representing the underlying reality) and using the Poisson estimate $\sigma_{i}^{2}=\lambda_{i}$ to capture the observational variance. Summing these realisations, one per bin over all magnitudes, provides a large set of plausible alternative totals. Figure~\ref{rates} shows histograms of these simulated totals for each of the four catalogues, fitted with Poisson distributions for reasons we now explain.

It is straightforward to show analytically that the sum of independent Poisson variables is itself Poisson with a mean (and hence variance) equal to the sum of the component means {\boldmath$\lambda$} \citep{johnson}. This result holds for (i) any number of independent Poisson variables (in the current context, bins) with (ii) any relationship {\boldmath$\lambda$}$=f(${\bf M}$)$, since the result is independent of $f(${\bf M}$)$. In the case of earthquakes placed into bins of width $\Delta M$ at magnitudes {\bf M}, for example, $f(${\bf M}$)$ is commonly fitted by a power-law or gamma distribution as in Fig.~\ref{fmd}. From the Poisson property $\sigma^{2}=\lambda$, it follows that
\begin{equation}
\sigma_{N}^{2}=\lambda_{N}=\sum\mbox{\boldmath$\lambda$}=\sum f\left({\bf M}\right)\label{eq4}.
\end{equation} 

Thus we have a useful result: if there exists a physically justifiable function that provides a satisfactory fit to the histogram (that is, Poisson-distributed uncorrelated residuals as in Fig.~\ref{resid}) then the mean and variance of the total number of events, over different realisations of the catalogue, are both equal to the sum of the fitted values (Eq.~\ref{eq4}). For the simulations of our four example catalogues (Fig.~\ref{rates}), we have mean total event numbers of $\lambda_{N}=19231,17491,46454,9301$ respectively; these match the actual observed totals to an accuracy of $\pm1$. Empircal evaluations confirm $\sigma_{N}=\sqrt{\lambda_{N}}$ to two significant figures, hence our estimated uncertainties on total event numbers for these catalogues are $\sigma_{N}=140,130,220,96$. Since (i) a Poisson distribution converges towards a Gaussian as $\lambda\rightarrow\infty$, (ii) a reasonable approximation to this exists where $\lambda>5$ and $S-\lambda>5$ for sample size $S$ \citep{leach}, and (iii) we have $S=10^{5}$ with $\lambda_{N}$ given above, it is not surprising that the Poisson confidence intervals for $\lambda_{N}\pm\sigma_{N}$ are (to two significant figures) 68\% as in the Gaussian case. 

\section{Conclusions}\label{conc} 

The purpose of this paper is to draw attention to the simplicity with which one can formally estimate event-rate uncertainties for applications in seismic hazard analysis, both in small magnitude ranges and over whole catalogues. For each of the four earthquake catalogues considered here, we find that the best estimate of both the mean and the variance of the total number of events, is equal to the total calculated from the fit to the histogram. This approximation holds where (i) the residuals of the fit are independently Poisson distributed, and (ii) the overall duration of the catalogue is greater than that of any significant correlations between either magnitudes or inter-event times. Note that the ratio of binomial-to-Poisson variance for any frequency $n$ is $\sigma^{2}_{b}/\sigma^{2}_{P}=1-p_{n}<1$, which implies that the Poisson approximation provides an upper bound for the uncertainty on the total event rate should any residuals generalise to the binomial case. However, correlations between inter-event times could cause significant future changes in event rates, greater than predicted by the naive estimates of uncertainty presented here, and this is the subject of further study. 

\section*{Acknowledgements}

The authors gratefully acknowledge financial support from the New and Emerging Science and Technologies Pathfinder program ``Triggering Instabilities in Materials and Geosystems'' (contract NEST-2005-PATH-COM-043386).

\end{document}